# Simultaneous three-wavelength unwrapping using external digital holographic multiplexing module

Nir A. Turko, Pinkie Jacob Eravuchira, Itay Barnea, Natan T. Shaked*

*Department of Biomedical Engineering, Faculty of Engineering, Tel-Aviv University, Tel-Aviv 69978, Israel*

*Corresponding author: nshaked@tau.ac.il

**We present an external interferometric setup that is able to simultaneously acquire three wavelengths of the same sample instance without scanning or multiple exposures. This setup projects onto the monochrome digital camera three off-axis holograms with rotated fringe orientations, each from a different wavelength channel, without overlap in the spatial-frequency domain, and thus allows the full reconstructions of the three complex wavefronts from the three wavelength channels. We use this new setup for three-wavelength phase unwrapping, allowing phase imaging of thicker objects than possible with a single wavelength, but without the increased level of noise that is typical to 2wl. We demonstrate the proposed techniques for metrological samples, including microchannel profiling and label-free cell imaging.**

http://dx.doi.org/10.1364/OL.99.099999

Digital holographic microscopy (DHM) has been a well-known modality in interferometric imaging over the last two decades [1]. This technique stands out for its ability to produce high-quality wide-field phase images with short acquisition times, which has made DHM a leading prospect, specifically in the field of biomedical imaging [2,3], where the microscopic samples are mostly transparent and pose most applicable data within the phase information. For objects that are optically thicker than the illumination wavelength used, DHM suffers from $2\pi$ phase ambiguities [1–4]. Therefore, the inherent range of DHM is limited to a single wavelength in optical path difference (OPD) units for transmission mode, or half of a wavelength for reflection mode.

The most common solution for this problem is digital one-wavelength phase unwrapping (1wl) [4], based on scanning the phase map and searching for $2\pi$ phase discontinuities. Whenever a discontinuity is found, the consecutive pixels are offset accordingly, so that the final phase is no longer limited to $-\pi$ to $\pi$. Most biological cells in vitro have rather smooth and mild-slope phase profiles, and thus algorithm-detected phase discontinuities reliably indicate on the corrections required. However, many metrological samples may have sharp phase profiles that differ by more than one wavelength between two spatial points, which will be incorrectly deciphered by the digital unwrapping algorithm [5]. This poses a major drawback for traditional DHM, which might be misleading, if relatively large steps are measured.

This phase unwrapping problem can be solved by using multi-wavelength DHM [5–11]. Shortly, two digital holograms of the sample are captured, each by a distinct wavelength, yielding two single-wavelength wrapped phase maps. The difference of both wrapped phase maps produces a new phase map of a much larger synthetic wavelength, which is increased when the difference between wavelengths decreases. Thus, the synthetic wavelength phase map is unwrapped optically. Dual wavelength DHM was therefore used for many applications in the optical metrology field, where objects, tens of μm thick, are examined, two orders of magnitude thicker than possible with single-wavelength DHM.

The main problem of two-wavelength phase unwrapping (2wl) is the inherent amplification of noise, resulting in decreased overall sensitivity by the same factor of two orders of magnitude in comparison to 1wl. In order to optically unwrap the phase map of thick objects, but still keep the sensitivity as that of a single wavelength, three-wavelength phase unwrapping (3wl) was introduced [12–15]. Using a third-wavelength hologram, one can produce phase maps of three gradually decreasing synthetic wavelengths, and then use hierarchical optical phase unwrapping [13]. However, the prospects of the 3wl techniques are encumbered by the fact that scanning the three wavelengths consecutively [14,15] makes this system also subject to vibrations and effectively slows down throughput by a factor of three for dynamic samples. On the other hand, if the three-wavelength holograms are acquired simultaneously, the overall system complexity dramatically increases, with the need for three co-located interferometers, containing three different reference arms, one per each wavelength, separated from the sample arm, which makes the system bulky and hard to align. Furthermore, such a system is highly subjective to sample movements and vibrations.

Lately, we proposed an external, portable, and nearly common-path interferometric module, which is capable of forming multiplexing of two wavelength holograms for 2wl [5]. The system was demonstrated as an efficient metrology tool on reflective pillars and step objects. However, as mentioned earlier, dual wavelength DHM does not keep the phase sensitivity of single wavelength DHM, typically in the range of a few nanometers.

In the current Letter, we propose a new external, nearly common path interferometric module for performing simultaneous 3wl using off-axis holographic multiplexing, based on integrating our dual-wavelength system [5] and our field of view tripling system [16]. The module projects on the monochrome digital camera three off-axis image holograms, each of a different wavelength and with a different off-axis fringe direction, preventing overlap of the three complex wavefronts in the spatial frequency domain of the multiplexed hologram, acquired by a single camera exposure. We thus obtain high sensitivity phase unwrapping for both reflection and transmission imaging modes, as experimentally demonstrated in this Letter.

Figure 1(a) introduces the proposed self-referencing, three-wavelength multiplexing DHM module that can be positioned at the exit port of a simple microscopic imaging system, which is illuminated by a coherent or partially coherent light source. In the interferometric module, lens L1 is positioned at a focal distance from the image plane produced by the microscope. Beam splitter BS splits the beam into a reference arm at the right and a sample arm at the top. The sample arm contains two consecutive 4f lens configurations (L1-L2, followed by L2-L4) for relay imaging upon the camera of all three wavelengths simultaneously. In the reference arm, a pinhole plate PH is located at the Fourier plane of L1, filtering spatial data of the sample and essentially creating the reference for all three wavelengths. The reference beams are then split by dichroic mirror DM1, where both shorter wavelengths are reflected to dichroic mirror DM2, after which each wavelength is separated. Mirrors M2, M3 and M4 are located at a focal distance from L3 in order to keep the 4f lens configuration, similarly to the sample arm. These mirrors are positioned in the image plane and each reflects its respective wavelength at a different angle, where L3 later focuses the wavelengths at three different regions on PH, as shown in Fig. 1(a) inset. Three off-axis holes are drilled into PH, passing the back-reflected beams. The reference beams are then recombined with the sample beams at BS, where lens L4 projects all three pairs of beams on the monochrome camera, where the multiplexed off-axis hologram is created.

We first illuminated the imaging system by wavelengths $\lambda_1 = 490$ nm, $\lambda_2 = 534$ nm, and $\lambda_3 = 692$ nm, originated by a supercontinuum source (Fianium SC-4), connected to acousto-optical tunable filter (Fianium AOTF) with a spectral bandwidth of around 5 nm. The imaging system used several microscope objectives depending on the samples imaged: 20×, 0.4 NA, long working distance; 20×, 0.42 NA, and long working distance 50×, 0.6 NA. The tube lens had a focal length of 180 mm. The focal lengths of the module lenses were f1 = 160 mm, f2 = 200 mm, f3 = 200 mm, and f4 = 200 mm. The cut-off wavelength of DM1 and DM2 were 610 nm and 498 nm, respectively. The pinhole diameter was 30 μm. The digital camera used was a monochrome CMOS camera (DCC1545M, Thorlabs) with 1280×1064 pixels of size 5.2 μm.

Figure 1(b) shows the resulting multiplexed off-axis image hologram, recorded by the monochrome camera in a single exposure. As shown in the magnified inset, this hologram contains off-axis fringes in three different orientations, depending on the angles induced by M2, M3 and M4. These angles can be selected such that the cross-correlation terms, resulted from each of the three holograms at each of the wavelength channels, $\lambda_1$, $\lambda_2$ or $\lambda_3$, will not overlap in the spatial frequency domain of the multiplexed hologram, as shown in Fig. 1(b), while using the camera spatial bandwidth more efficiently. Note that since we use three wavelength channels, no interference between them is created so that we do not get any cross-talks between the channels in the spatial frequency domain. The captured wavefront at each wavelength can then be obtained by cropping the relevant cross-correlation term and inverse-Fourier transforming it. The argument of the exponent in each of the resulting complex wavefront matrices is the coinciding wrapped phase map in each wavelength channel [5].

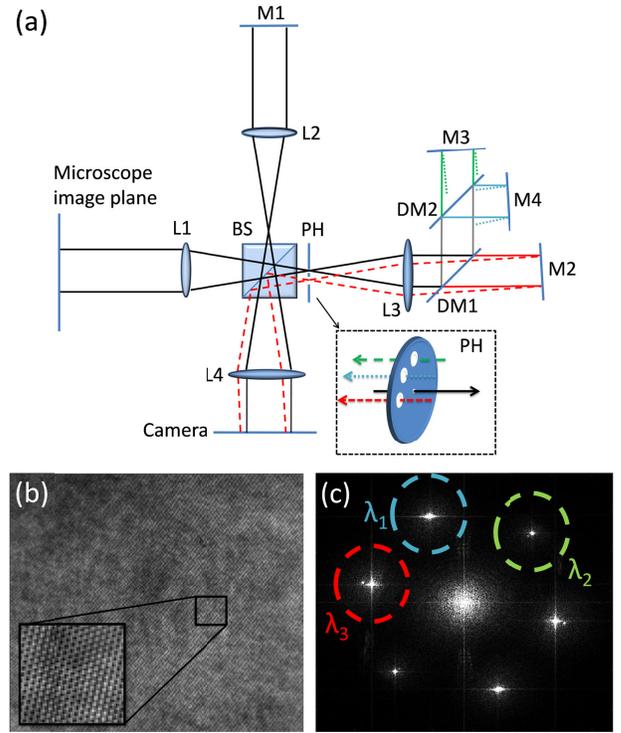

Fig. 1. (a) The proposed three-wavelength interferometric module. L1-L4, achromatic lenses. BS, beam-splitter. M1-M4, mirrors. PH, pinhole plate, with three additional off-axis holes drilled. Camera, monochrome digital camera. Back-reflected beams from M3 and M4 are not drawn, for simplicity. The inset at the bottom shows a perspective view of PH that spatial filters the light for the incident beam, creating a reference beam, whereas the three wavelength channels are reflected from mirrors M2-M4 and pass through the holes drilled in PH. (b) Multiplexed hologram recorded by the camera in a single exposure, containing off-axis fringes at three different orientations. (c) The coinciding power spectrum, consisting of the auto-correlation terms at the centers, and three distinct cross-correlation term pairs from the three wavelength channels.

Since the acquired phase is inversely related to the wavelength, the $2\pi$ jumps will occur in different locations in the three phase maps, where each map bounds to measure OPD within the unambiguous range of its wavelength. In multi-wavelength optical unwrapping, we use a higher synthetic wavelength $\Lambda_{1-2} = \lambda_1 \lambda_2 / (\lambda_1 - \lambda_2)$ to create a new phase map as follows:

$$\varphi_{\Lambda_{1-2}} = \varphi_1 - \varphi_2 = \frac{2\pi}{\Lambda_{1-2}} \cdot \Delta n \cdot h \qquad (1)$$

where OPD = $\Delta n \cdot h$ is the product of the refractive index difference $\Delta n$ and the sample thickness $h$. This procedure allows increasing the measurement range while decreasing the sensitivity by the same factor. Hierarchical optical unwrapping [13] allows maintaining both the increased measurement range and the original phase sensitivity. Using the phase map of the third wavelength, it is possible to obtain six different phase maps, with decreasing effective wavelengths. Each larger-wavelength phase map is used for unwrapping the data of the slightly lower-wavelength phase map, as follows. First, all phase maps $\varphi_{\Lambda 1-2}$, $\varphi_{\Lambda 2-3}$, $\varphi_{\Lambda 1-3}$ are computed according to Eq. 1. The algorithm starts with the unwrapped phase map acquired by the largest synthetic wavelength. The wrapped phase map (represented by an apostrophe) of the next greatest synthetic wavelength is then subtracted from the former step unwrapped map and the integer of the difference is calculated:

$$N = floor\left( \frac{\varphi_{\Lambda 1-2} - \varphi'_{\Lambda 2-3}}{\Lambda_{2-3}} + 0.5 \right) \quad (2)$$

The integer $N$ represents how many multiplications of the shorter wavelength must be added to the wrapped phase map, in order to get the unwrapped phase map of the next step as follows:

$$\varphi_{\Lambda 2-3} = \varphi'_{\Lambda 2-3} + \Lambda_{2-3} \cdot N. \quad (3)$$

The result is an unwrapped phase map, corresponding to the shorter wavelength, with lower noise level compared to the phase map obtained with 2wl. This process may be repeated with lower wavelengths until the required sensitivity is achieved. The smaller synthetic wavelength has to exceed the phase noise of the larger synthetic wavelength phase data [13].

Since three phase maps are required for hierarchical phase unwrapping, it is crucial to record their coinciding holograms as fast as possible, to eliminate environmental noise factors originated from the sample dynamics or the stability of the system. As presented above, the interferometric module proposed provides the ability to acquire all three holograms simultaneously.

Figure 2 shows a relatively shallow biological cell (from WM-115 malignant melanoma cell line), imaged with the proposed module, with $\lambda_1$ = 490 nm, $\lambda_2$ = 534 nm and $\lambda_3$ = 692 nm. The wrapped phase of a single wavelength $\lambda_1$ is shown in Fig. 2(a), and its digitally unwrapped version is shown in Fig. 2(b). However, the two-dimensional digital unwrapping is computationally expensive due to searching for the best gradients of adding $2\pi$ to correct for phase ambiguities, which might limit certain applications such as real-time visualization [17]. Multi-wavelength phase unwrapping processing is significantly faster, as it relies solely on several summations and multiplications of matrices. As can be seen in Fig. 2(c) and (d), presenting the phase maps of $\varphi_{\Lambda 1-2}$, $\varphi_{\Lambda 2-3}$ obtained simultaneously with the proposed module, 2wl dramatically increases the spatial noise of the coinciding phase maps, due to the increase of the respective synthetic wavelength. Figure 2(e) presents the unwrapped phase map obtained by 3wl. With the exception of a few pixels due to slight inconsistencies between the wavelengths, this phase map is similar to the one presented in Fig. 2(b), obtained with the digital and much slower 1wl procedure.

Next, we examined the proposed module for profiling transparent micro-channels. For this, we fabricated 7 - µm - high PDMS micro channels in a cross shape. The $\Delta n$ for the wavelengths used were [18]: 0.422 for $\lambda_1$ = 490 nm, 0.418 for $\lambda_2$ = 534 nm, and 0.415 for $\lambda_3$ = 692 nm. The first synthetic wavelength was $\Lambda_{1-2}$ = 5.95 µm, where the actual OPD was 2.8 µm, within the unambiguous range. Figure 3 shows the phase maps of various stages of the hierarchical phase unwrapping process. First, in Fig. 3(a), the wrapped phase map of a single wavelength $\lambda_1$ is shown. The height is wrapped between $-\lambda_1/2$ to $\lambda_1/2$ and the standard deviation of noise level is 16.3 nm at the middle of the channel intersection. Figure 3(b) shows the phase map $\varphi_{\Lambda 1-2}$, resulting from 2wl. While we now cover a larger height range, the noise level has also increased and now stands at 213.8 nm at the same spatial position. Figure 3(c) shows the next stage of the unwrapping process, at the next hierarchical level of $\Lambda_{2-3}$, where the range remains the same as on the first stage, but the noise levels decreases to 61.6 nm. The final stage of the process is presented in Fig. 3(d), where both the large height range of 2wl and the low noise level of 1wl are retained.

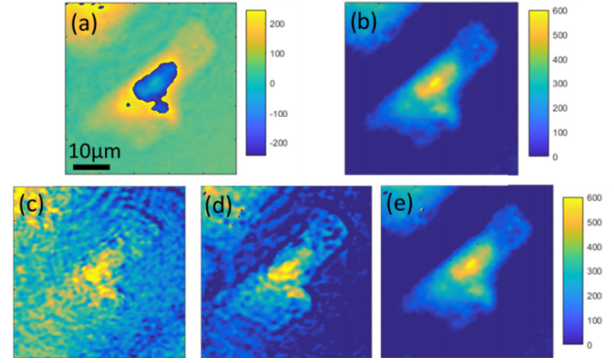

Fig. 2. (a) Wrapped phase map of a biological cell. (b) Unwrapped phase map obtained by applying conventional, computationally expensive 1wl algorithm. (c) Two wavelength unwrapped phase map, $\varphi_{\Lambda 1-2}$. (d) Two wavelength unwrapped phase map, $\varphi_{\Lambda 2-3}$. (e) Three wavelength unwrapped phase map. Colorbar represents OPD in nm.

Cross-sections around two 150-200-nm-high manufacturing defects in the micro-channel, indicated by the red arrows in Fig. 3(a), are presented in Fig. 4(a) (defect across upper dashed line) and Fig. 4(b) (lower defect). 1wl can profile these defects well, since they are within the ambiguous range. However, 2wl missed the defects due to noise increase on the entire phase map. On the other hand, 3wl successfully profiled the defects, demonstrating the importance of keeping the measurement height sensitivity of 1wl. Figure 4(c) shows the height cross-section across the channel at the dashed line in Fig. 3(d) along all 4 stages of the hierarchical phase unwrapping. As the unwrapping progresses, the noise levels decrease, while still keeping a large unambiguous range.

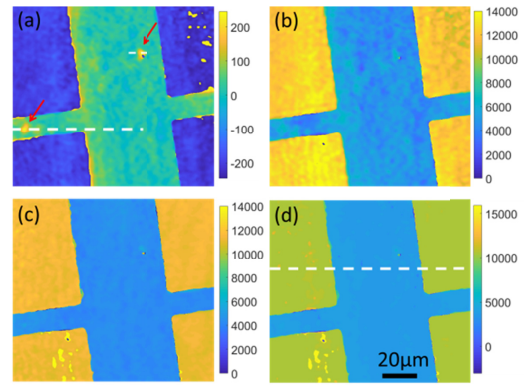

Fig. 3. (a) Unwrapped phase map obtained by 1wl. The height difference detected is erroneous. Two lithography defects are indicated by red arrows. (b) The coinciding phase map $\varphi_{\Lambda 1-2}$ obtained by 2wl, (c) The coinciding phase map obtained by 3wl during the first stage of the process. (d) The coinciding phase map obtained by the final stage of 3wl. Colorbar represents height in nm.

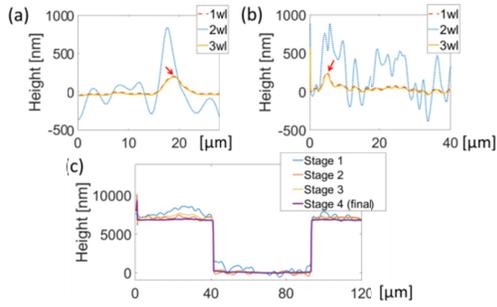

Fig. 4. (a,b) Cross sections around the defects, across the upper and lower dashed lines, respectively, shown in Fig. 3(a), compared to the coinciding cross-sections obtained by 2wl and 3wl. (c) Cross sections across the dashed line shown in Fig. 3(d), for all stages of the hierarchical unwrapping.

To test the robustness of the proposed module, we changed the imaging setup to a reflection microscope, and measured a reflective 10.8-μm-high silicon step target, created by etching and verified by a white-light interferometry commercial instrument (Contour GT, Bruker). To enable imaging such a deep sample, we chose closer wavelengths, $\lambda_1$ = 490 nm, $\lambda_2$ = 607 nm and $\lambda_3$ = 622 nm, resulting in synthetic wavelengths of $\Lambda_{2-3}$ = 25.170 μm, $\Lambda_{1-2}$ = 2.542 μm and $\Lambda_{1-3}$ = 2.309 μm. It should be noted that while the module itself did not require any modifications when switching from transmission to reflection illumination modes, the processing must be adapted to consider the double-pass of the beam when it interacts with the reflective sample, effectively halving the synthetic wavelength. This allows an unambiguous range of 12.585 μm when using $\Lambda_{1-2}$, which is still suitable for the step target measured. Figure 5(a) shows a three-dimensional rendering of the resulting step target topography. Figure 5(b) shows a histogram of the height values at the top and bottom of the step. As can be seen, the height estimation using 2wl is much less certain than 3wl, due to the noise amplification in 2wl.

Spatial standard deviation measurements were taken for the 1wl, 2wl and 3wl results, at the same area of 24 × 24 μm². For $\lambda_1$ = 490 nm, the standard deviation was 2.5 nm, for $\Lambda_{1-2}$ = 2.542 μm, it was 12.3 nm, and for the largest wavelength, $\Lambda_{2-3}$ = 25.170 μm, it was 182.1 nm. As expected, the noise increased by about the same factor as the wavelengths. On the other hand, the standard deviation in 3wl was 2.5 nm, the same as 1wl. Thus, using 3wl hierarchical algorithm, the data is far more precise, yielding a more robust and reliable measurement for metrology.

To conclude, we presented a self-referencing three-wavelength DHM module for optical unwrapping of either thick transparent or high reflective samples. We provided several experimental demonstrations for the module flexibility and reliability by varying the wavelengths, microscope objectives and illumination modes. Note that the module uses hologram multiplexing while utilizing the same number of camera pixels and sharing its dynamic range. The main advantage of this 3wl method is its capability to maintain high-grade phase sensitivity (with standard deviation of 2.5 nm), while measuring the 3-D profile of samples of up to 10.8 μm with accuracy of 50 nm. The proposed module requires the illumination to be highly coherent, or with low temporal coherence still providing off-axis interference on the entire sensor area [3] and has wavelength regions that are fully separable. Using a single sample beam and splitting the reference beam makes all wavelength channels easy to align and obsoletes spatial coherence issues. Furthermore, the overall processing time required deceases significantly, since a 2-D digital unwrapping process of 1wl can be more than six times slower than the two IFFT [3] added to the 3wl process. The proposed module can work The use of a multiplexed self-referencing module for the acquisition of all three holograms in a single exposure makes it less susceptible to sample vibrations and environmental noises, making it a potentially useful tool for many applications, including high-throughput DHM in metrology, where acquisition speed and a large unambiguous range are of great importance.

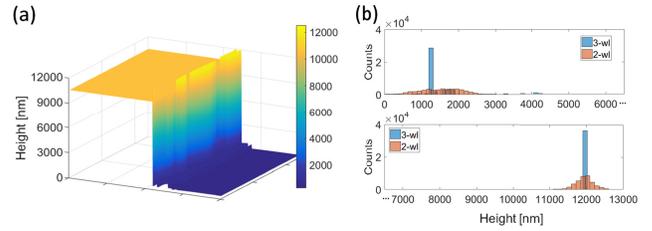

Fig. 5. (a) A 3-D rendering of the height map of a reflecting step target measured by the proposed module. (b) The coinciding histograms of different locations on the uniform step, when unwrapped by two (orange) and three (blue) wavelengths, demonstrating significantly lower noise for 3wl.

**Funding.** H2020 European Research Council (ERC) (678316).

**References with titles**